\documentclass[preprint,aps,tightenlines,showpacs,superscriptaddress]{revtex4}
\usepackage[dvips]{graphicx}
\usepackage{bm}
\usepackage{dcolumn}
\usepackage{epsfig}

\begin{document}
\title{Double and single pion photoproduction within a dynamical coupled-channels model }
\author{H. Kamano}
\affiliation{Excited Baryon Analysis Center (EBAC), Thomas Jefferson National
Accelerator Facility, Newport News, Virginia 23606, USA}
\author{B. Juli\'a-D\'{\i}az}
\affiliation{Department d'Estructura i Constituents de la Mat\`{e}ria
and Institut de Ci\`{e}ncies del Cosmos,
Universitat de Barcelona, E--08028 Barcelona, Spain}
\affiliation{Excited Baryon Analysis Center (EBAC), Thomas Jefferson National
Accelerator Facility, Newport News, Virginia 23606, USA}
\author{T.-S. H. Lee}
\affiliation{Excited Baryon Analysis Center (EBAC), Thomas Jefferson National
Accelerator Facility, Newport News, Virginia 23606, USA}
\affiliation{Physics Division, Argonne National Laboratory,
Argonne, Illinois 60439, USA}
\author{A. Matsuyama}
\affiliation{Department of Physics, Shizuoka University, Shizuoka 422-8529, Japan}
\affiliation{Excited Baryon Analysis Center (EBAC), Thomas Jefferson National
Accelerator Facility, Newport News, Virginia 23606, USA}
\author{T. Sato}
\affiliation{Department of Physics, Osaka University, Toyonaka,
Osaka 560-0043, Japan}
\affiliation{Excited Baryon Analysis Center (EBAC), Thomas Jefferson National
Accelerator Facility, Newport News, Virginia 23606, USA}
 
\begin{abstract}
Within a dynamical coupled-channels model that has already been fixed by
analyzing the data of the $\pi N \to \pi N$ and $\gamma N \to \pi N$ reactions,
we present the predicted double pion
photoproduction cross sections 
up to the second resonance region, $W< 1.7$ GeV. The 
roles played by the different mechanisms within our model
in determining both the
single and double pion photoproduction reactions are analyzed, focusing on
the effects attributable to the direct $\gamma N\to\pi\pi N$ mechanism, the
interplay between the
resonant and nonresonant amplitudes, and the  coupled-channels effects.
The model parameters that
can be determined most effectively in the combined studies of both the
single and double pion photoproduction data are identified 
for future studies.

\end{abstract}

\pacs{13.75.Gx, 13.60.Le, 14.20.Gk}

\maketitle

\section{Introduction}
The spectrum and structure of low-lying nucleon and $\Delta$ 
resonances (collectively referred as $N^*$)  are 
 primordial information for any understanding of the nonperturbative 
QCD domain. Consequently, a great effort has been made at the
Excited Baryon Analysis Center (EBAC) during the past few
years to extract the properties of $N^*$ from the world data on 
$\pi N \to \pi N$ and $\gamma N \to \pi N$ data~\cite{bl04}.

It is well acknowledged nowadays that a proper extraction and further 
interpretation of $N^*$ properties require the construction of 
reaction models that maintain the unitarity of most relevant channels
and can correlate the vast amount of data for both the single 
and double meson production reactions. Among the 
existing theoretical approaches, the one taken at EBAC 
tries to encompass the aforementioned by 
considering the interactions among
the  $\gamma N$, $\pi N$, $\eta N$, and  $\pi \pi N$  channels 
within a multichannel, multiresonance 
framework~\cite{msl07}. After constraining the hadronic part of the 
model by fitting~\cite{jlms07} the $\pi N \rightarrow \pi N$ scattering data,
 we have performed 
our first studies of single pion photoproduction~\cite{jlmss08} 
and electroproduction reactions~\cite{jklmss09}.

As discussed in our previous works, the hadronic part of the model was
constrained mostly using $\pi N \to \pi N$ experimental data. This means 
that the couplings of the $N^*$ to the $\pi \Delta$, $\rho N$ and 
$\sigma N$ channels, which are the quasi-two-body channels
of the $\pi\pi N$, are necessarily not well constrained in the 
current version of the model.
 To this extent, double pion photoproduction 
reactions are important for understanding the way $N^*$ couple to the 
$\pi \pi N$ channel, and thus to refine our global dynamical 
coupled-channels framework. In Ref.~\cite{kjlms09}, we
carried out such a study  for 
$\pi N \rightarrow \pi \pi N$ reactions with the predicted 
cross sections in reasonable agreement with
the available data. In this work, we extend that work to
investigate double pion photoproduction reactions by comparing
our predictions with the total cross sections 
data~\cite{a68,b95,wo00,la01,a03,ah03,ah05} and invariant mass 
distributions~\cite{wo00,la01,be04}.
We first  present the predictions of our model for the
double pion photoproduction reactions up to $W=1.7$ GeV.
We then analyze how the discrepancies with the data are sensitive to
which of the electromagnetic parameters of the model, as a step toward 
performing the combined fits of the world data on $\pi N, \gamma N \rightarrow \pi N, \pi\pi N$ reactions.
 
Most of the previous investigations of the double pion 
photoproduction reactions employed the tree-diagram 
models~\cite{gomez95,ochi97,nacher00,fix05},
 emphasized the 
roles of certain resonances on specific double pion photoproduction 
reactions, or focused on the very near threshold 
region using chiral perturbation theory~\cite{ch1,ch2}.
In our approach, we do not make such simplifications. We perform the
full coupled-channels calculations and include 
all channels and $N^*$ states determined in Refs.~\cite{jlms07,jlmss08}.

The basic formulas used in this work are presented in Sec.~\ref{sec:formula}. 
In Sec.~\ref{sec:direct} we present the predictions of the current model and
analyze   the contributions  from the direct $\gamma N \rightarrow \pi\pi N$
mechanism and the transitions from $\gamma  N$ to the
unstable $\pi\Delta$, $\sigma N$ and $\rho N$ states. 
In Sec.~\ref{sec:bare} we scrutinize the contribution 
of each of the $\gamma N \rightarrow N^*$ helicity amplitudes
on both single pion and double 
pion photoproduction reactions. A summary and some conclusions are given 
in Sec.~\ref{sec:summary}.

\section{Basic formulas}
\label{sec:formula}

\begin{figure}[t]
\centering
\includegraphics[clip,width=0.75\textwidth]{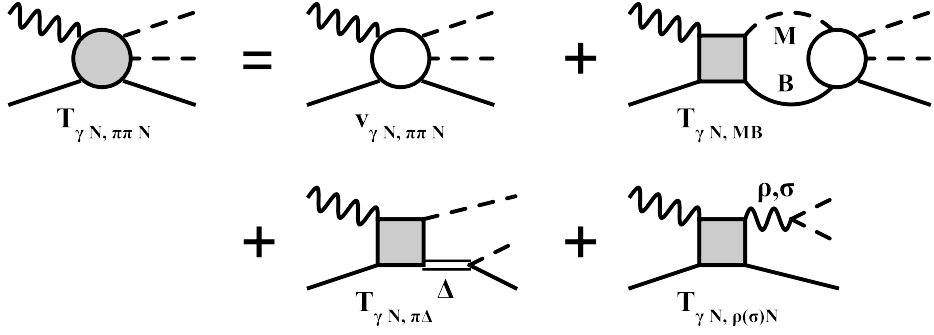}
\caption{Graphical representations of $T_{\gamma N,\pi\pi N}$
of Eqs.~(\ref{eq:tpipin-1})-(\ref{eq:tpipin-sigman}).}
\label{tg2p}
\end{figure}

\begin{figure}[b]
\centering
\includegraphics[clip,width=0.8\textwidth]{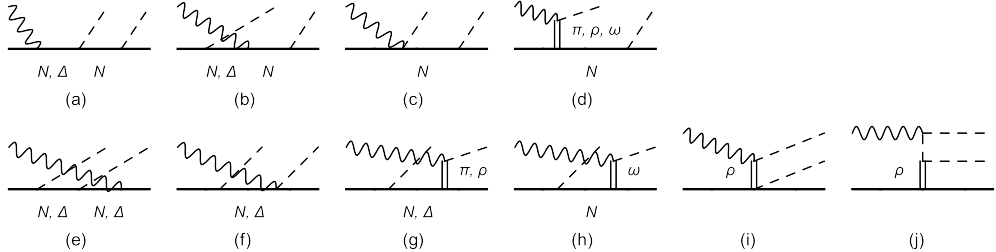}
\caption{Diagrams considered for $v_{\gamma N, \pi \pi N}$.
}
\label{vpot}
\end{figure}

Within the EBAC dynamical coupled-channels (EBAC-DCC) model, the $\gamma N\to
\pi\pi N$ amplitude consists of four pieces~\cite{msl07} 
(see Fig.~\ref{tg2p}):

\begin{eqnarray}
T_{\gamma N,\pi \pi N}(E) &=&
 T^{\text{dir}}_{\gamma N,\pi\pi N}(E)
+T^{\pi\Delta}_{\gamma N,\pi \pi N}(E)
+T^{\rho N}_{\gamma N,\pi\pi N}(E)
+T^{\sigma N}_{\gamma N, \pi \pi N}(E),
\label{eq:tpipin-1}
\end{eqnarray}
with
\begin{eqnarray}
T^{\text{dir}}_{\gamma N,\pi \pi N}(E)
=
 v_{\gamma N,\pi \pi N} +
\sum_{MB} T_{\gamma N,MB}(E)G_{MB}(E) v_{MB,\pi\pi N}
\,,
\label{eq:tpipin-dir} 
\end{eqnarray}
with, 
\begin{eqnarray}
T^{\pi \Delta}_{\gamma N,\pi\pi N}(E)
&=&
T_{\gamma N, \pi \Delta}(E)
G_{\pi\Delta}(E)
\Gamma_{\Delta \rightarrow \pi N}
\label{eq:tpipin-pid} , \\
& & \nonumber \\
T^{\rho N}_{\gamma N, \pi\pi N}(E)
&=& 
T_{\gamma N,\rho N}(E) G_{\rho N}(E) 
h_{\rho\rightarrow \pi\pi} 
\label{eq:tpipin-rhon},  \\
& & \nonumber \\
T^{\sigma N}_{\gamma N,\pi \pi N}(E)
&=&
T_{\gamma N,\sigma N}(E) G_{\sigma N}(E)
h_{\sigma\rightarrow \pi\pi}.
\label{eq:tpipin-sigman}
\end{eqnarray}
Here $\Gamma_{\Delta\rightarrow \pi N}$, $h_{\rho\rightarrow \pi\pi}$, and
$h_{\sigma\rightarrow \pi\pi}$  describe the  
$\Delta\to \pi N$, $\rho\rightarrow \pi\pi$, and 
$\sigma \rightarrow \pi\pi$ decays, respectively; $G_{MB}(E)$ 
($MB=\pi N,\eta N,\pi\Delta, \rho N, \sigma N$) 
are the meson-baryon Green's functions.
$v_{\gamma N, \pi\pi N}$ represents the direct $\gamma N\to\pi\pi N$ transition
potentials illustrated in Fig.~\ref{vpot}. 
The processes described by $v_{\gamma N, \pi\pi N}$ 
are not contained in the $T^{MB}_{\gamma N,\pi\pi N}$, and thus
there is no double counting.

The $\gamma N \rightarrow MB$ transition amplitudes 
can be divided into the so-called nonresonant and resonant 
amplitudes (suppressing angular momentum, isospin, and 
momentum indices),
\begin{eqnarray}
T_{\gamma N,MB}(E)  &=&  t_{\gamma N,MB}(E)
+ t^R_{\gamma N,MB}(E),
\label{eq:tmbmb}
\end{eqnarray}
with
\begin{eqnarray}
t_{\gamma N, MB}(E)
&=&
v_{\gamma N, MB}
+\sum_{M'B'}v_{\gamma N,M'B'} 
G_{M'B'}(E) t_{M'B',MB}(E).
\label{eq:pw-nonr}
\end{eqnarray}
and
\begin{eqnarray}
t^R_{\gamma N,MB}(E) &=&
 \sum_{N^*_i, N^*_j}
\bar{\Gamma}_{\gamma N\to N^*_i}(E)
[D(E)]_{i,j}
\bar{\Gamma}_{N^*_j\to MB}(E)\,.
\label{eq:pw-r}
\end{eqnarray}
In Eq.~(\ref{eq:pw-nonr}),
$v_{\gamma N, MB}$ represents the $\gamma N \rightarrow MB$ transition
potential derived from tree diagrams of a set
of phenomenological Lagrangians describing the interactions among
$\gamma$, $\pi$, $\eta$, $\rho$, $\omega$, $\sigma$, $N$, and
$\Delta$(1232) fields. The details are given explicitly in Appendix F
of Ref.~\cite{msl07}. The dressed $\gamma N \rightarrow N^*$ vertex function 
appearing in Eq.~(\ref{eq:pw-r}) is defined by
\begin{eqnarray}
\bar{\Gamma}_{\gamma N\to N^*}(E)
&=&
{\Gamma}_{\gamma N \to N^*}
+
\sum_{M'B'} v_{\gamma N,M'B'} G_{M'B'}(E)
\bar{\Gamma}_{M'B'\to N^*}(E),
\label{eq:pw-v}
\end{eqnarray}
where ${\Gamma}_{\gamma N \to N^*}$ denotes the bare 
$\gamma N \rightarrow N^*$ vertex within the EBAC-DCC model
and is parametrized as
\begin{eqnarray}
{\Gamma}^{J}_{N^*,\lambda_\gamma\lambda_N}(q)
& =&\frac{1}{(2\pi)^{3/2}}\sqrt{\frac{m_N}{E_N(q)}}\frac{1}{\sqrt{2q}}
[\sqrt{2q_R} A^{J}_{\lambda}]
\delta_{\lambda, (\lambda_\gamma-\lambda_N)} \,,
\label{eq:ggn}
\end{eqnarray}
where $q_R$ is defined by the $N^*$ mass $M_{N^*} = q_R+E_N(q_R)$.

Within our model, the meson-baryon Green function $G_{MB}$,
the hadronic nonresonant amplitude $t_{MB,M'B'}$,
the dressed $N^\ast$ propagator $D(E)$,
and 
the dressed $N^\ast\to MB$ vertex function $\bar \Gamma_{N^\ast\to MB}$
are purely hadronic processes.
We take these hadronic pieces from
the model constructed from analyzing the data of
$\pi N \rightarrow \pi N$ scattering~\cite{jlms07},
and keep them fixed throughout this paper.

The calculation of the terms $T^{MB}_{\gamma N\to\pi\pi N}$ with
$MB=\pi\Delta,\rho N,\sigma N$, defined by
Eqs.~(\ref{eq:tpipin-pid})-(\ref{eq:tpipin-sigman}), 
is straightforward. However, the calculation of the second term of
$T^{\text{dir}}_{\gamma N\to\pi\pi N}$, defined by Eq.~(\ref{eq:tpipin-dir}), 
is much more complex. 
To simplify the calculation, we employ the 
same prescription as in the calculation of the $\pi N\to\pi\pi N$ 
reactions~\cite{kjlms09}.
This is based on the observation that the processes illustrated in
Figs.~\ref{vpot}(a)-\ref{vpot}(d) 
can be written as
\begin{eqnarray}
v_{\gamma N,\pi\pi N}^{(\text{a-d})} &\sim&v_{\gamma N,\pi N}G_{\pi N}(E)h_{N\to\pi N},
\end{eqnarray}
where $v_{\gamma N,\pi\pi N}^{(\text{a-d})}$ is the sum of the all processes 
illustrated in Figs.~\ref{vpot}(a)-\ref{vpot}(d),
$v_{\gamma N,\pi N}$ is the two-body $\gamma N\to \pi N$ potential,
and $h_{N\to\pi N}$ is the $N\to\pi N$ vertex function.
Taking account of only a part of $v_{MB,\pi\pi N}$
that can be approximately expressed as
$v_{MB,\pi\pi N}\sim v_{MB,\pi N}G_{\pi N}(E)h_{N\to\pi N}$,
Eq.~(\ref{eq:tpipin-dir}) can be written as
\begin{eqnarray}
T^{\text{dir}}_{\gamma N,\pi \pi N}(E) &\sim&
v_{\gamma N,\pi\pi N}^{(\text{e-j})}+
[
v_{\gamma N,\pi N} +\sum_{MB}T_{\gamma N,MB}(E)
\;
G_{MB}(E) v_{MB,\pi N}
]
G_{\pi N}(E)h_{N\to\pi N}
\nonumber\\
&=&
v_{\gamma N,\pi\pi N}^{(\text{e-j})}
+T_{\gamma N,\pi N}G_{\pi N}(E)h_{N\to \pi N}.
\nonumber\\
&&
\label{eq:tpipin-dir2}
\end{eqnarray}
Here in the last step we have used the relation
$T_{\gamma N,\pi N} = v_{\gamma N,\pi N} 
+ \sum_{MB}T_{\gamma N,MB}G_{MB}v_{MB,\pi N}$.
We use Eq.~(\ref{eq:tpipin-dir2}) which can be calculated with all parameters
taken from our previous analysis of $\pi N, \gamma N \rightarrow \pi N$
reactions.

The formulas for calculating total cross sections and 
invariant mass distributions
from our amplitudes can be found in Ref.~\cite{kjlms09} and
are not shown here.

\section{Analysis of the direct reaction mechanisms and the coupled-channels effect}
\label{sec:direct}
With the parameters determined from our previous analysis of
$\pi N, \gamma N\to \pi N$ reactions~\cite{jlms07,jlmss08},
the results presented in this section are 
pure predictions within the current model developed in EBAC.
We first present our results of the double pion photoproduction reactions, 
and then examine how the reactions mechanisms within our model
determine the cross sections.

\begin{figure}[t]
\centering
\includegraphics[clip,width=0.8\textwidth]{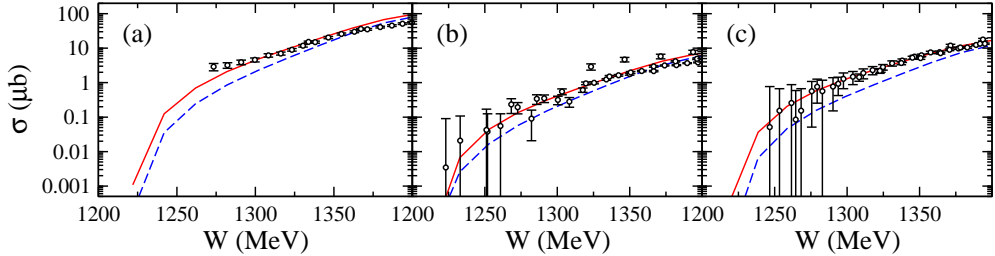}
\caption{Near threshold behavior of the total cross section for
$\gamma p \to  \pi \pi N$:
(a) $\gamma p\to\pi^+\pi^-p$,
(b) $\gamma p\to\pi^0\pi^0p$,
and (c) $\gamma p\to\pi^+\pi^0n$.
The red solid curve is the full results predicted from our current model, 
and the blue dashed curves
are the results without the $T^{\text{dir}}_{\gamma N,\pi\pi N}$ contribution.
The data are taken from Refs.~\cite{a68,b95,wo00,la01,a03,ah03,ah05}.
}
\label{fig:g2ptcs-th}
\end{figure}

\begin{figure}[t]
\centering
\includegraphics[clip,width=0.67\textwidth]{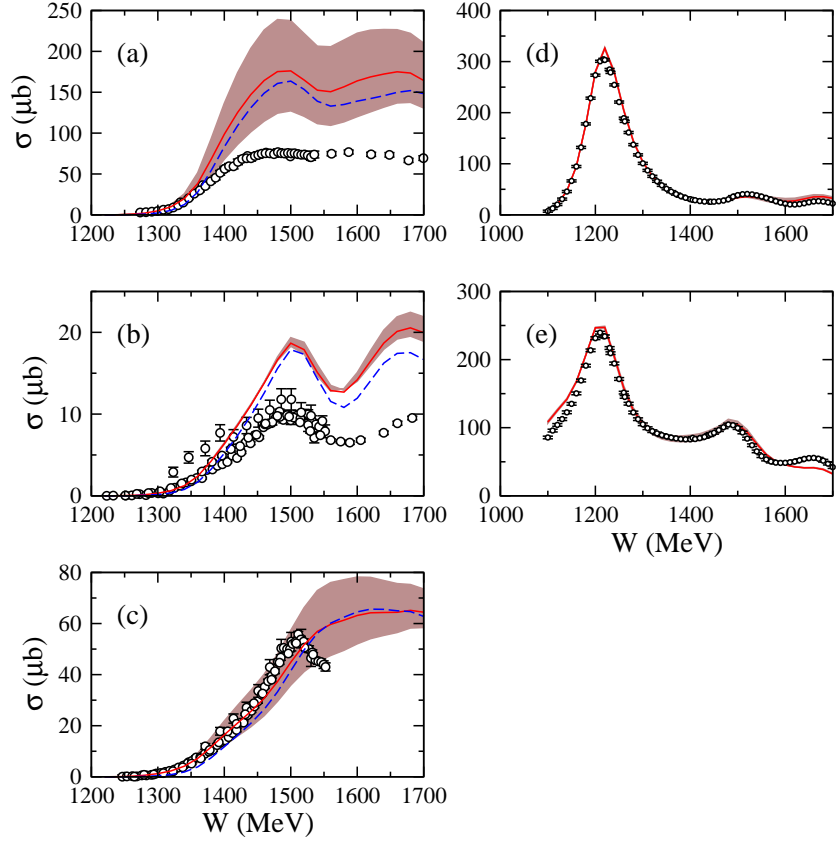}
\caption{Total cross sections of the double and single pion photoproduction
reactions up to $W=1.7$ GeV: 
(a) $\gamma p \to \pi^+\pi^-p$, 
(b) $\gamma p \to \pi^0\pi^0p$, (c) $\gamma p \to \pi^+\pi^0n$,
(d) $\gamma p \to \pi^0 p$, and (e) $\gamma p \to \pi^+n$. 
The red solid curve is the full result predicted from our current 
model, and the blue dashed curve in (a)-(c) is the result
without $T^{\text{dir}}_{\gamma N,\pi\pi N}$ contribution. The band is generated 
by allowing a $25\%$ variation in the value of the $\pi N\Delta$ coupling
constant $g_{\pi N \Delta}$ used in the electromagnetic amplitudes. 
The data of the double and single pion photoproduction reactions 
are taken from Refs.~\cite{a68,b95,wo00,la01,a03,ah03,ah05}
and Refs.~\cite{saidweb}, respectively.
}
\label{fig:g2p-g1p-tcs}
\end{figure}

In Fig.~\ref{fig:g2ptcs-th}, we find that our current
model (red solid curve) has a good agreement with the $\gamma N\to\pi\pi N$ 
total cross sections in the energy region up to $W=1.4$ GeV.
We observe that the direct $T^{\text{dir}}_{\gamma N,\pi\pi N}$
amplitude can greatly improve the model to reproduce the near threshold
behavior of the $\gamma N\to \pi\pi N$ 
total cross section data.
Its effects in  higher $W$ are shown in
Figs.~\ref{fig:g2p-g1p-tcs}(a), \ref{fig:g2p-g1p-tcs}(b), 
and~\ref{fig:g2p-g1p-tcs}(c).
The red solid curves are the predictions from our full calculations 
and the blue dashed curves are from turning off 
the term $T^{\text{dir}}_{\gamma N,\pi\pi N}$ (the
bands in the figure will be explained later in this paper). 
We see that the effect of $T^{\text{dir}}_{\gamma N,\pi\pi N}$ is 
sizable on $\gamma p \rightarrow \pi^+\pi^- p$ [Fig.~\ref{fig:g2p-g1p-tcs}(a)] 
and $\gamma p \rightarrow \pi^0\pi^0 p$ [Fig.~\ref{fig:g2p-g1p-tcs}(b)], and negligible 
on $\gamma p \rightarrow \pi^+\pi^0 n$ [Fig.~\ref{fig:g2p-g1p-tcs}(c)].
It is  clear that its inclusion does not change the 
energy dependence of the total cross sections for any of the considered 
$\gamma N \to \pi \pi N$ reactions.

Although the threshold behavior is in general well reproduced 
as can be seen in Fig.~\ref{fig:g2ptcs-th}, our predictions at higher $W$
shown
in Fig.~\ref{fig:g2p-g1p-tcs}
clearly overestimate the experimental data above $W=1.4$ GeV 
in both $\gamma p \to \pi^+ \pi^- p$ and 
$\gamma p \to \pi^0 \pi^0 p$ 
reactions, while the results of $\gamma p \to \pi^+ \pi^0 n$ are 
good up to $W=1.5$ GeV. 
However, our current model reproduces the 
$\gamma N\to\pi N$ reactions quite well in the considered 
energy region, as seen in the right panels of Fig.~\ref{fig:g2p-g1p-tcs}.
This fact indicates that there exist reaction processes which 
have significant effect on the observables of $\gamma N\to \pi\pi N$, 
but not of $\gamma N\to\pi N$.

\begin{figure}[t]
\centering
\includegraphics[clip,width=0.75\textwidth]{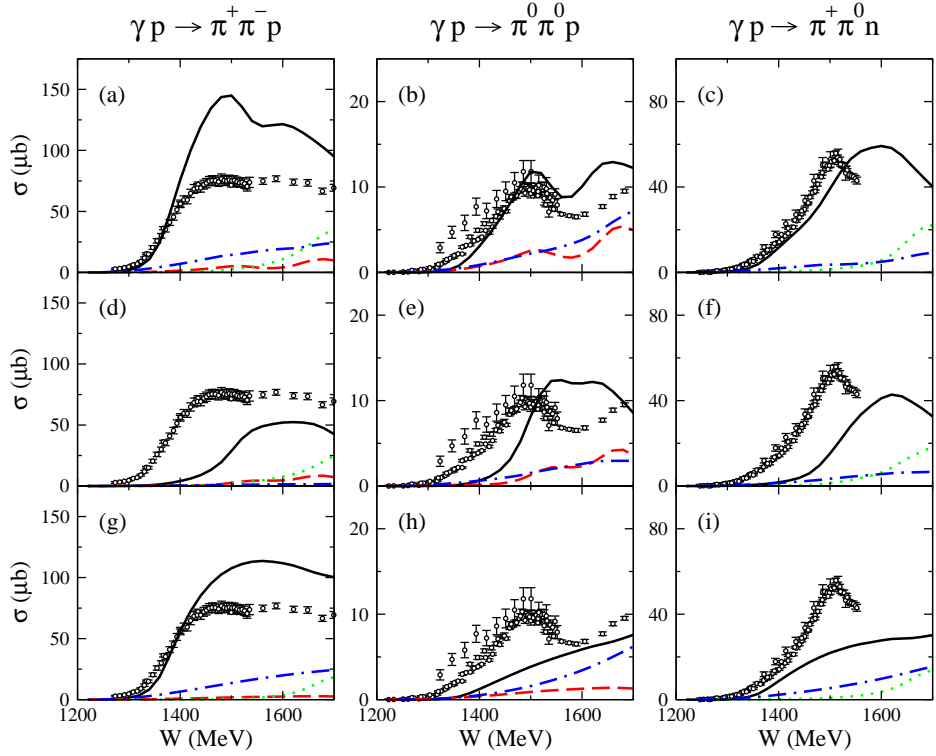}
\caption{
Contributions of each reaction process described in
Eqs.~(\ref{eq:tpipin-dir})-(\ref{eq:tpipin-sigman})
to the total cross sections.
(Black solid) $\gamma N\to\pi\Delta$ contribution 
($T^{\pi\Delta}_{\gamma N,\pi\pi N}$); 
(red dashed)$\gamma N\to\sigma N$ contribution
($T^{\sigma N}_{\gamma N,\pi\pi N}$); 
(green dotted)$\gamma N\to\rho N$ contribution
($T^{\rho N}_{\gamma N,\pi\pi N}$); 
(blue dashed-dotted) the direct contribution
($T^{\text{dir}}_{\gamma N,\pi\pi N}$).
(Top row) Full results of each contribution; 
(Middle row) Results with the replacement of
$T_{\gamma N,MB}\to t^R_{\gamma N,MB}$;
(Bottom row) Results with the replacement of
$T_{\gamma N,MB}\to t_{\gamma N,MB}$.
The data are taken from Refs.~\cite{a68,b95,wo00,la01,a03,ah03,ah05}.
}
\label{fig:g2ptcs-each}
\end{figure}

To get some insights into our disagreement with the data and to guide our 
future combined analysis of all 
$\pi N, \gamma N \rightarrow \pi N, \pi\pi N$ reactions,
 we examine which mechanisms 
are most relevant to our calculations in 
this energy region. 
We first examine the contributions of each
process appearing in Eqs.~(\ref{eq:tpipin-dir})-(\ref{eq:tpipin-sigman}).
The results from the full amplitude are shown in the top row of
Fig.~\ref{fig:g2ptcs-each}:
$T^{\pi\Delta}_{\gamma N,\pi\pi N}$ (black solid),
$T^{\sigma N}_{\gamma N,\pi\pi N}$ (red dashed), 
$T^{\rho N}_{\gamma N,\pi\pi N}$ (green dotted), and
$T^{\text{dir}}_{\gamma N,\pi\pi N}$ (blue dash-dotted).
The figures in the left, middle, and right columns
are of the $\gamma p\to\pi^+\pi^- p$, $\gamma p\to\pi^0\pi^0 p$, and
$\gamma p\to\pi^+\pi^0 n$ total cross sections, respectively.

We also show in the middle (bottom) row 
of Fig.~\ref{fig:g2ptcs-each}
the results for which the full two-body 
amplitude $T_{\gamma N,MB}$ in 
Eqs.~(\ref{eq:tpipin-pid})-(\ref{eq:tpipin-sigman}) and~(\ref{eq:tpipin-dir2})
is replaced with its resonant (nonresonant) part
$T_{\gamma N,MB}\to t^R_{\gamma N,MB}$ 
($T_{\gamma N,MB}\to t_{\gamma N,MB}$). 
Thus we can examine the relative 
importance between different mechanisms in resonant 
$t^R_{\gamma N,MB}$ and nonresonant $t_{\gamma N,MB}$
amplitudes separately.
Note that the curves describing the $\gamma N\to\sigma N$ ($\gamma N\to\rho N$)
process are not seen in
the $\gamma p\to\pi^+\pi^0 n$ ($\gamma p\to\pi^0\pi^0 p$) total cross sections
because the corresponding terms do not  contribute because of isospin
selection rules.
In Figs.~\ref{fig:g2ptcs-each}(a)-\ref{fig:g2ptcs-each}(c),
we clearly see that the full $\gamma N\to\pi\Delta\to\pi\pi N$ 
processes (black solid curves) 
have the largest contribution compared to the other 
processes. By comparing Figs.~\ref{fig:g2ptcs-each}(a) and~\ref{fig:g2ptcs-each}(g), we
further find that the large discrepancy with the $\gamma p\to\pi^+\pi^- p$
data 
is attributable mainly to the nonresonant $\gamma N\to\pi\Delta\to\pi\pi N$ 
amplitude.
The dominance of the nonresonant $\gamma N\to\pi\Delta\to\pi\pi N$ in
all three $\gamma N\to\pi\pi N$ reactions can 
also be seen  in the bottom panels of  Fig.~\ref{fig:g2ptcs-each}. 

Most of the nonresonant $\gamma N \to \pi \Delta$ transition matrix elements
considered in our model depend on the $\pi N \Delta$ coupling constant
$g_{\pi N\Delta}$ (see Ref.~\cite{msl07} for the details). 
We thus examine how our predictions are sensitive to this coupling strength.
This is illustrated in Fig.~\ref{fig:g2p-g1p-tcs} where we have presented
bands, which are generated by varying $g_{\pi N \Delta}$ included in the 
$\gamma N \to \pi \Delta$ transition matrix elements by $\pm 25 \%$. 
Clearly  such changes in $g_{\pi N \Delta}$ have a great influence on
$\gamma p\to\pi^+\pi^-p$ (top) and $\gamma p\to\pi^+\pi^0 n$ (bottom), and 
less of an influence on $\gamma p\to\pi^0\pi^0 p$ (middle).
Within our dynamical coupled-channels model,
the $\gamma N\to\pi\Delta$ process also enters 
in the single pion photoproduction reactions as a consequence of
the unitarity, and thus its change
consistently affects the single pion photoproduction observables, too.
As can be seen in the right panels of Fig~\ref{fig:g2p-g1p-tcs},
its importance turns out to be
very minor in the $\gamma N\to\pi N$ total cross sections. The bands from 
varying $g_{\pi N \Delta}$ in $\gamma N \rightarrow \pi\Delta$ by
$\pm 25 \%$ are not visible.
From this observation, in the remainder of this paper 
we will use a 20\% smaller value for the $g_{\pi N\Delta}$
appearing in the electromagnetic potentials. 
The value turns out to be very close to that of the quark model.

\begin{figure}[t]
\centering
\includegraphics[clip,width=0.75\textwidth]{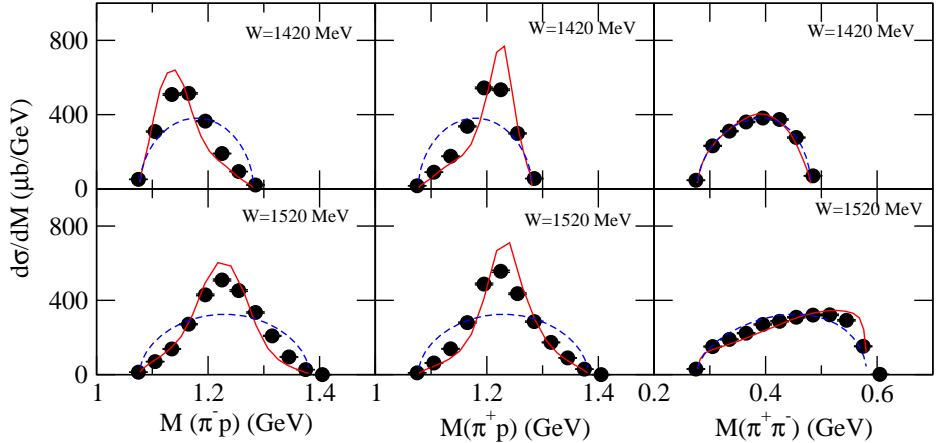}
\caption{
Invariant mass distributions of $\gamma p \to \pi^+ \pi^- p$
at $W=1420,1520$ MeV:
(left) $(\pi^- p)$; (middle) $(\pi^+ p)$; (right) $(\pi^+ \pi^-)$.
The red solid curve is the full result, and
the blue dashed curve is the phase space distribution.
The magnitude of both curves is normalized to the data.
The data are taken from Ref.~\cite{be04}.
}
\label{fig:invms-pm}
\end{figure}

\begin{figure}[t]
\centering
\includegraphics[clip,width=0.5\textwidth]{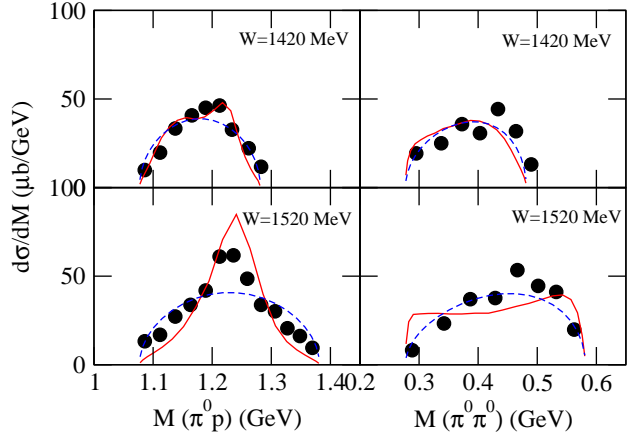}
\caption{
Invariant mass distributions of $\gamma p \to \pi^0 \pi^0 p$
at $W=1420,1520$ MeV:
(left) $(\pi^0 p)$; (right) $(\pi^0 \pi^0)$. 
The red solid curve is the full result, and
the blue dashed curve is the phase space distribution.
The magnitude of both curves is normalized to the data.
The data are taken from Ref.~\cite{wo00}.
The energy bins of the data are $20$-$30$ MeV around the central $W$ shown in the panels.
}
\label{fig:invms-00}
\end{figure}

\begin{figure}[t]
\centering
\includegraphics[clip,width=0.75\textwidth]{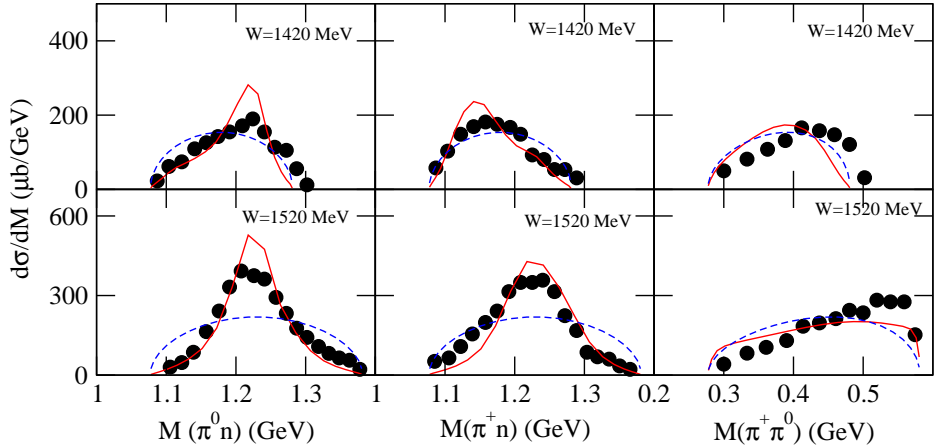}
\caption{Invariant mass distributions of $\gamma p \to \pi^+ \pi^0 n$
at $W=1420,1520$ MeV:
(left) $(\pi^0 n)$; (middle) $(\pi^+ n)$; (right) $(\pi^+ \pi^0)$.
The red solid curve is the full result, and
the blue dashed curve is the phase space distributions.
The magnitude of both curves is normalized to the data.
The data are taken from Ref.~\cite{la01}.
The energy bins of the data are $20$-$30$ MeV around the central $W$ shown in the panels.
}
\label{fig:invms-p0}
\end{figure}

In Figs.~\ref{fig:invms-pm}-\ref{fig:invms-p0},
we show the predicted invariant 
mass distributions of
$\gamma p \to \pi^+ \pi^- p$, 
$\gamma p \to \pi^0 \pi^0 p$, and
$\gamma p \to \pi^+ \pi^0 n$, respectively. 
To compare with the shapes 
of the data, the overall magnitudes 
of our predictions (red solid curves) are normalized 
to have the same integrated values of the data.
We can see that the shapes of the predicted 
$\pi N$ invariant mass distributions 
are in reasonable agreement with the data
for all cases considered, while
deviations are seen in several $\pi\pi$ invariant mass
distributions
(right panels of  Figs.~\ref{fig:invms-pm}-\ref{fig:invms-p0}).

This is found to be attributable to the fact that
the $\pi N$ distributions are dominated by
the $\Delta(1232)$ in the $\gamma N \rightarrow \pi\Delta(1232)\to \pi \pi N$
process,
while the $\pi\pi$ distributions involve the interferences
among all of the
$\gamma N\to \pi\Delta, \rho N, \sigma N \to \pi\pi N$ amplitudes.
The results of the $\pi\pi$ invariant mass distributions 
have provided useful information for improving our current model.
In particular, the deviations from the data in 
the $\pi^+\pi^0$ distributions of the $\pi^+\pi^0 n$ channel 
at high invariant mass (right panels of Fig.\ref{fig:invms-p0})
suggest that the parameters associated with the $\rho N$ channels 
will need to be modified.

\begin{figure}[t]
\centering
\includegraphics[clip,width=0.75\textwidth]{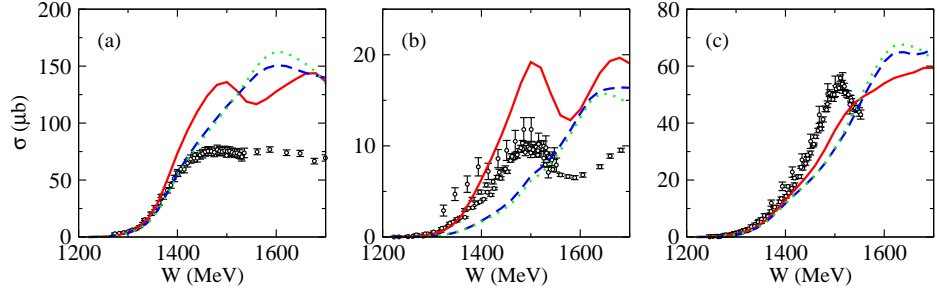}
\caption{
Coupled-channels effects associated with electromagnetic interactions.
The red solid curve is full results, 
the green dotted curve corresponds to taking only 
the diagonal element in the $M'B'$ summation
in Eqs.~(\ref{eq:pw-nonr}) and~(\ref{eq:pw-v}), 
and the blue dashed curve is obtained by further making 
a replacement of
$t_{\gamma N,MB}\to v_{\gamma N,MB}$ and 
$\bar \Gamma_{\gamma N\to N^\ast} \to \Gamma_{\gamma N\to N^\ast}$.
The data are taken from Refs.~\cite{a68,b95,wo00,la01,a03,ah03,ah05}.
}
\label{fig:c4}
\end{figure}

The most relevant novelty of the present study is the use of a dynamical
coupled-channels model. 
In Fig.~\ref{fig:c4}, we show the coupled-channels effects
associated with the electromagnetic interactions on
the $\gamma N\to\pi\pi N$ total cross sections,
which is demonstrated here for the first time in the investigations
of double pion photoproduction reactions.
The red solid curves are our full results.
The green dotted curves are the results in which
only the diagonal part ($M'B' = MB$) is taken in the $M'B'$ summation
of Eqs.~(\ref{eq:pw-nonr}) and~(\ref{eq:pw-v}),
and the blue dashed curves are obtained by further setting
$t_{\gamma N,MB}\to v_{\gamma N,MB}$ 
and 
$\bar \Gamma_{\gamma N\to N^\ast}\to \Gamma_{\gamma N\to N^\ast}$. 
These correspond to examining the coupled-channels effect
associated with the electromagnetic interactions.
(Note again that the pure hadronic part of the amplitudes
is fixed with the model determined in Ref.~\cite{jlms07} 
throughout this paper.)
In the considered energy region up to $W=1.7$ GeV,
we find that the blue dashed and green dotted curves almost overlap
with each other but both of them are quite different
from our full results (red solid curves).
This suggests that the structure in the 
$\gamma p \to \pi^+\pi^- p,~\pi^0\pi^0p$ total cross sections
is attributable mostly to the couplings between reaction channels.

Before closing this section, we comment on the recent measurements
of the polarization observables.
It was shown in Refs.~\cite{st05,kr09}
that existing reaction models have significant discrepancies
in the beam-helicity asymmetry measured
at CLAS~\cite{st05} and more recently at MAMI~\cite{kr09}.
We have observed that our current model also produces similar
discrepancies to that of other works shown in Refs.~\cite{st05,kr09}.
These results indicate that the polarization observables will provide critical
information on constraining reaction models and understanding the $N^\ast$ states.

\section{Effects of resonances }
\label{sec:bare}

\begin{table}[b]
\centering
\caption{The bare $\gamma N \rightarrow N^*$ helicity amplitudes determined 
from $\chi^2$-fits to the $\gamma N \rightarrow \pi N$.
The asterisks in the second (third) column mark
the $N^*$ states in which $\gamma N$ transition process is found to be 
relevant to the single (double) pion photoproduction reactions 
up to $W=1.7$ GeV.}
\label{tab1}
\ruledtabular
\begin{tabular}{cccrr} 
Bare  $N^*$   &$\gamma N\to\pi N$& $\gamma N\to\pi\pi N$ & $A_{1/2} [10^{-3}$ GeV$^{-1/2}]$ & $A_{3/2} [10^{-3}$ GeV$^{-1/2}]$      \\ \hline
$S_{11} (1535)$&* &* &   100& ---         \\
$S_{11} (1650)$& &  &$-$19& ---        \\ 
$S_{31} (1620)$&* &* &  203& ---       \\ 
$P_{11} (1440)$& & &$-$17& ---        \\ 
$P_{13} (1720)$& &  &$-53$&$-$21        \\
$P_{33} (1232)$&* & &$-$78&$-$129        \\ 
$D_{13} (1520)$&* &* &   44&$-$60     \\ 
$D_{15} (1675)$& &  &   54&   30    \\ 
$D_{33} (1700)$& & & 0.3 &$-$64    \\ 
$F_{15} (1680)$&* & &$-$82&$-$69   \\ 
\end{tabular}
\end{table}

The bare helicity amplitudes, defined in Eq.~(\ref{eq:ggn}), are free 
parameters in our framework. They quantify the photoexcitation of the 
core $N^*$ states and, together with their dressed counterparts, are 
to be interpreted by means of microscopic models (e.g., quark models, 
lattice QCD calculations). Although $A_{\lambda}^J$ are taken to be real 
numbers, the dressed helicity amplitudes, which have in general a 
sizable contribution from the second term in Eq.~(\ref{eq:pw-v}), are 
complex numbers. This second term contains the meson-cloud contribution 
to the $\gamma N N^*$ vertex, which is to a large extent fixed from 
the strong interaction sector.

In this section we present the effect on the single and double pion 
photoproduction observables of variations on the bare helicity 
amplitudes, which affect directly the dressed ones, see Eq.~(\ref{eq:pw-v}).
This will be done by presenting results computed by varying the 
initial value of the bare helicity amplitudes
listed in Table~\ref{tab1}, by $\pm 50\%$. 
The results are presented as bands 
in the figures, these bands are generated by filling the region enclosed 
by curves from two calculations using $0.5 \times A_{j/2}$ 
and $1.5 \times A_{j/2}$.

Before proceeding to showing our results, 
we comment on the bare helicity
amplitudes presented in Table~\ref{tab1}.
Those values are not exactly the same as those of
our previous $\gamma N\to\pi N$ analysis~\cite{jlmss08}.
There we did not provide any measure of the uncertainty in the 
bare helicity amplitudes that resulted from fitting 
the photoproduction data. In the current paper 
we have varied the binning of the data and thus some of the 
less constrained helicities resulting from the fit are 
varied. In the following we will quantify the effect of 
such variations, providing a clear indicator of the 
dependence of our results for both single and double pion 
production on the helicity amplitudes. 

\subsection{S-wave $N^*$s}

\begin{figure}[t]
\centering
\includegraphics[clip,width=0.75\textwidth]{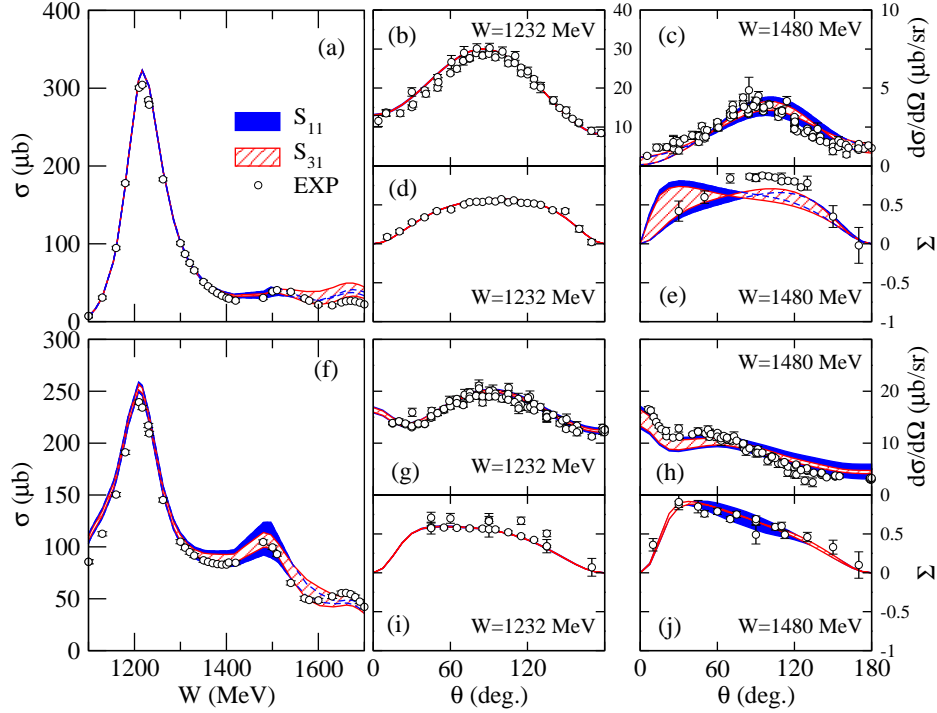}
\caption{The panels (a)-(e) depict the total cross section, 
differential cross sections and 
photon asymmetry
for $\gamma p \to \pi^0p$, 
and (f)-(j) show total cross section, differential 
cross sections and 
photon asymmetry
for $\gamma p \to \pi^+n$. 
Each band is obtained by allowing a $50\%$ variation 
of the helicity amplitudes for 
the $A_{1/2}$ of $S_{11}(1535)$ (solid blue) and 
$A_{1/2}$ of $S_{31}(1620)$ (oblique-lined red) listed in Table~\ref{tab1}.
The data are taken from Ref.~\cite{saidweb}.}
\label{fig:c1}
\end{figure}

We start the comparison with the $S_{11}(1535)$ and $S_{31}$(1620). 
In Fig.~\ref{fig:c1} we show the effect of varying their helicity 
amplitudes on the single pion photoproduction data. The sample data we 
consider are the total cross sections (left panels)
for $\gamma p \to \pi^0p$ and 
$\gamma p \to \pi^+n$ and differential cross sections and polarization 
data in the $\Delta(1232)$ region (middle panels)
and in the $W=1500$ MeV region (right panels).

First we note that the $\pm 50\%$ change in helicity amplitudes
for the $S_{11}(1535)$ resonance plays an important 
role in building the peak near the $1500$ MeV region for
both $\gamma p \to \pi^+n, \pi^0p$ 
total cross sections[see Figs.~\ref{fig:c1}(a) and~\ref{fig:c1}(f)] 
and correspondingly in the differential cross section near the 
$1500$ MeV region [see Figs.~\ref{fig:c1}(c) and~\ref{fig:c1}(h)]. 
The $S_{31}$ gives a 
prominent contribution in the whole energy region above the $\Delta$ (1232)
region, as indicated by the oblique-lined bands.
The $S_{31}$ also affects the 
forward peaking of the $\gamma p \to \pi^+n$ differential cross 
section data around $W=1500$ MeV [see Fig.~\ref{fig:c1}(h)]. Their influence 
on the 
photon asymmetry $\Sigma$
is sizable and qualitatively similar for 
both resonances, being negligible in the $\Delta(1232)$ region.

\begin{figure}[h]
\centering
\includegraphics[clip,width=0.85\textwidth]{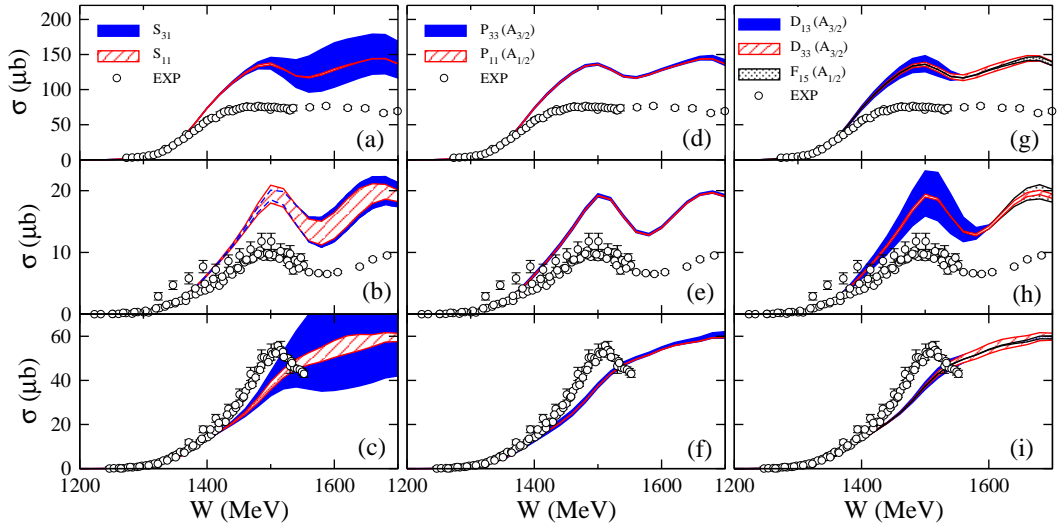}
\caption{Total cross sections, [panels (a), (d), and (g)] 
$\gamma p \to \pi^+\pi^-p$, 
[panels (b), (e), and (h)] 
$\gamma p \to \pi^0\pi^0p$, and 
[panels (c), (f), and (i)] $\gamma p \to \pi^+\pi^0n$. 
The different bands are generated by allowing $\pm 50\%$ variations 
of the helicity amplitudes listed in Table~\ref{tab1}.  
The data are taken from Refs.~\cite{a68,b95,wo00,la01,a03,ah03,ah05}.
}
\label{fig:res}
\end{figure}
Now we turn to the double pion photoproduction reactions, see left column of 
Fig.~\ref{fig:res}. First, as expected, and the same occurs 
for all resonances considered, the helicities have no 
influence on the near threshold behavior. Second, both S-wave 
resonances play a relevant role for the considered reactions. 
Modifying the $A_{1/2}$ of the $S_{11}$(1535), the total cross 
sections for $\gamma p \to \pi^0\pi^0 p$ and 
$\gamma p \to \pi^+ \pi^0 n$ can vary up to $20\%$, although 
there is no qualitative change in the energy dependence of 
the total cross sections [see Figs.~\ref{fig:res}(b) and~\ref{fig:res}(c)]. 
The $S_{31}$ case is similar, but actually 
affects all the reactions. 
A smaller value of the $S_{31}$ helicity amplitude
is suggested by these results. 
Within our model, none of the peaks seen in the total cross 
section data can be ascribed solely to S-wave resonances. 

\subsection{P-wave $N^*$s}

\begin{figure}[t]
\centering
\includegraphics[clip,width=0.75\textwidth]{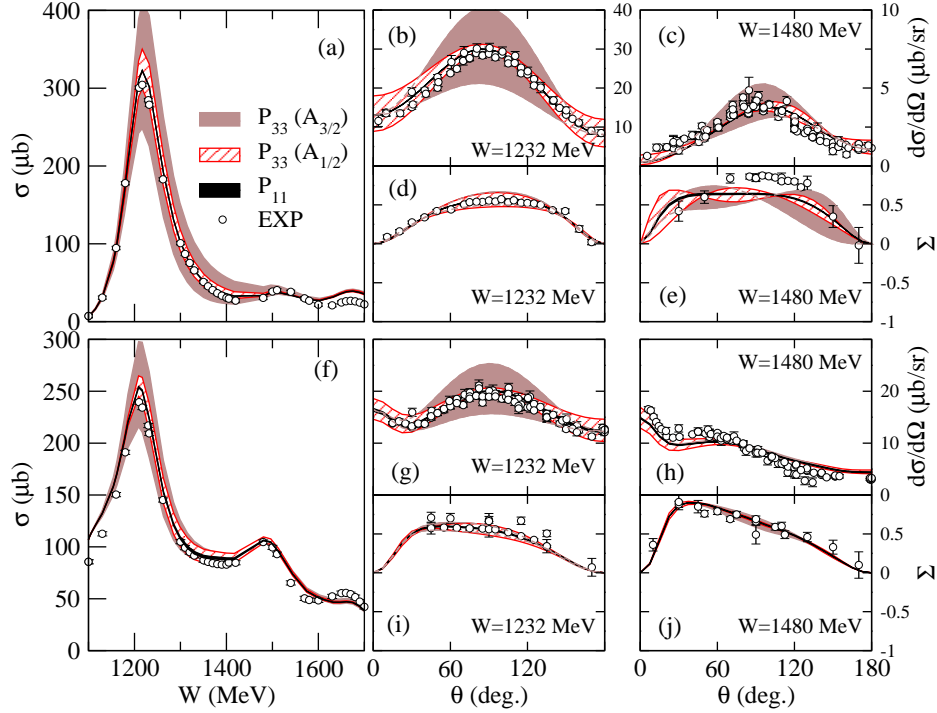}
\caption{The panels (a)-(e) depict the total cross section, differential 
cross sections and 
photon asymmetry
for $\gamma p \to \pi^0p$, 
and (f)-(j) show total cross section, differential 
cross sections and 
photon asymmetry for $\gamma p \to \pi^+n$. 
Each band is obtained by allowing a $50\%$ 
variation of the helicity amplitudes for the 
$A_{3/2}$ of $P_{33}(1232)$ (solid brown),
$A_{1/2}$ of $P_{33}(1232)$ (oblique-lined red), and
$A_{1/2}$ of $P_{11}(1440)$ (solid black)
listed in Table~\ref{tab1}.
The data are taken from Ref.~\cite{saidweb}.
}
\label{fig:c2}
\end{figure}

The helicity amplitudes of the $\Delta$(1232) resonance are essentially 
fixed by analyzing data near its nominal mass, as has long 
been known. In Fig.~\ref{fig:c2} we fully confirm this. The 
effect of variations on both $A_{1/2}$ and $A_{3/2}$ of the 
$\Delta(1232)$ is well localized around its peak but reaches 
up to 300 MeV above it in the $\gamma p \to \pi^0p$ reaction
[see Figs.~\ref{fig:c2}(a)-\ref{fig:c2}(e)]. 
This can also be seen in their influence on the photon
asymmetry at $W=1480$ MeV. The $A_{3/2}$ mostly affects 
the perpendicular angles, while the $A_{1/2}$ affects the 
forward and backward angles. 
The Roper resonance plays a minor role, with no sizable 
trace in the observables.

In the double pion photoproduction case, however (see middle column of 
Fig.~\ref{fig:res}), the $\gamma N$ transition processes of both
$P_{33}$ and $P_{11}$ play almost no 
role in the entire considered region. 
Let us point out that 
we refer here to the influence of the $P_{33}$ as an s-channel 
exchange, the importance of the $\Delta$ in this reaction is 
of course large, as pointed out in Sec.~\ref{sec:direct}, where 
we show that most of the reaction flows through the $\pi \Delta$ 
channel.

\subsection{D and F-wave $N^*$s}

\begin{figure}[t]
\centering
\includegraphics[clip,width=0.75\textwidth]{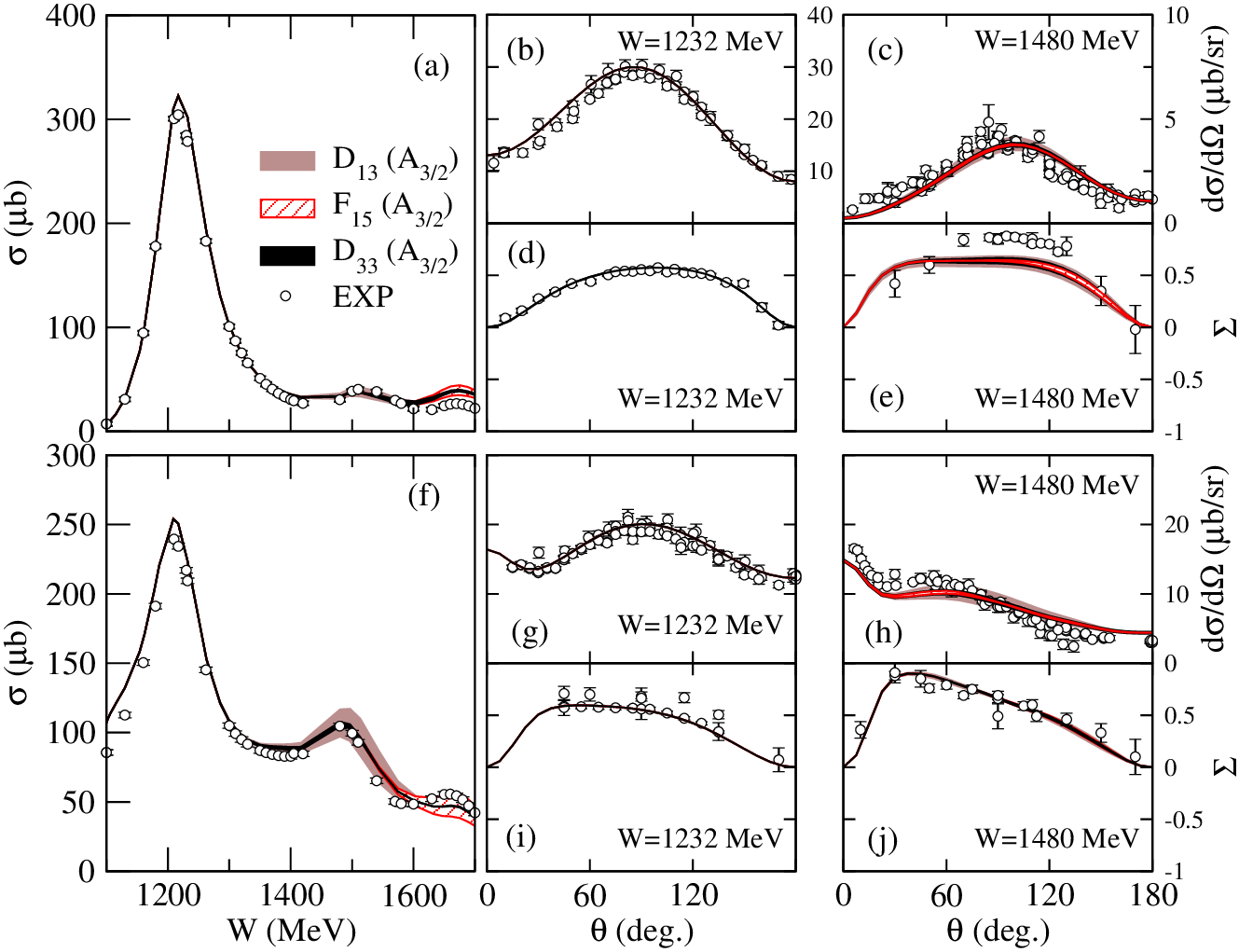}
\caption{The panels (a)-(e) depict the total cross section, differential 
cross sections and 
photon asymmetry 
for $\gamma p \to \pi^0p$, 
and (f)-(j) show total cross section, differential 
cross sections and 
photon asymmetry
for $\gamma p \to \pi^+n$. 
Each band is obtained by allowing a $50\%$ variation 
of the helicity amplitudes for the 
$A_{3/2}$ of $D_{13}(1520)$ (solid brown), 
$A_{3/2}$ of $F_{15}(1680)$ (oblique-lined red), and
$A_{3/2}$ of $D_{33}(1700)$ (solid black)
listed in Table~\ref{tab1}.
The data are taken from Ref.~\cite{saidweb}.
}
\label{fig:c3}
\end{figure}

Let us first study the influence of the helicity amplitudes on the 
single pion photoproduction. The $D_{13}$ is responsible 
for part of the second peak near 1500 MeV in the $\gamma p \to \pi^+ n$ 
total cross sections [see Fig.~\ref{fig:c3}(f)]. 
The $F_{15}$(1680) contributes to 
the third peak in both total cross sections. In the
middle and right panels of  Fig.~\ref{fig:c3}, we see that none 
of the $\pm 50\%$ changes of $D_{13}$, $F_{15}$ and $D_{33}$ 
helicity amplitudes affect much the 
$\Sigma$
and $d\sigma/d\Omega$ observables.

$D$ wave resonances have long been advocated as being responsible 
for most of the structure observed in the total cross sections 
for $\gamma p \to \pi \pi N$.  The first peak in the total cross 
sections has been explained in tree level calculations thanks 
to the $D_{13}$(1520)~\cite{ochi97,gomez95,fix05} and to interferences 
with the $D_{33}(1700)$~\cite{nacher00}. In our coupled-channels model 
we confirm the very important role played by the $D_{13}$(1520),
which builds up a large fraction of the first peak in the 
$\gamma p \to \pi^0\pi^0 p$ reaction [see the right panels in
Fig.~\ref{fig:res}]. On the other hand its effect is also sizable on 
the $\gamma p \to \pi^+\pi^-p$ total cross section, producing an 
overprediction of this observable in our model. As in the 
tree-diagram models of Refs.~\cite{ochi97,gomez95,fix05} the peak structure in 
this reaction is always much more pronounced in the models than 
in the experimental data. Effects of the $D_{33}$ are sizable only
on the $\gamma p \to \pi^+ \pi^0 n$, similar to what was reported in
Ref.~\cite{nacher00}, but they do not produce a peak structure as the 
experimental data show.

\section{Summary and conclusions}
\label{sec:summary}

Within the dynamical coupled-channels model constructed from
analyzing the single pion production reactions~\cite{jlms07,jlmss08}, 
we have investigated 
 the total cross sections and the invariant mass distributions 
for the double pion photoproduction reactions 
off the proton in the energy region up to $W=1.7$ GeV. 
In the low-energy region up to $W=1.4$ GeV,
our results agree well with the total cross sections data,
in which the direct process $T^{\text{dir}}_{\gamma N,\pi\pi N}$
plays a crucial role for the reproduction of the data.
Above $W=1.4$ GeV, our current model starts 
to overestimate the data for $\gamma p \to \pi^+\pi^-p$ and 
$\gamma p \to \pi^0\pi^0p$. 
We have found that the $\gamma N\to\pi \Delta$ process
is most relevant for the $\gamma N\to\pi\pi N$ reactions and
is a major origin of the overestimation
in the $\gamma p\to\pi^+\pi^- p$ total cross section. 
Our model reproduces well the shapes of
the invariant mass distributions data 
except for several $\pi\pi$ invariant mass distributions
of $\gamma p\to\pi^0\pi^0 p$ and $\gamma p\to\pi^+\pi^0 n$.
We expect that this deviation provides useful information
to improve our current model.
Also, we have demonstrated the coupled-channels effects on the double pion 
photoproduction case, which is of similar size to the $\pi N \to \pi \pi N$ 
case.

It is noted that our current model describes the single pion 
photoproduction observables in the same energy region quite well. 
We thus have examined the origins of our disagreements with the data
by considering both the single and the double photoproduction reactions.
We have found that the $\pi N\Delta$ coupling constant $g_{\pi N\Delta}$
in the $\gamma N \to MB$ transition matrix element
plays an important role.
If we reduce its strength determined in Ref.\cite{jlms07}
by  25 \% to a value close to the quark model value,
the magnitude of the $\gamma p \to\pi^+\pi^- p$ total cross section
is drastically reduced,
while the corresponding changes in the single pion photoproduction
observables are negligible.
This finding indicates that a smaller value of $g_{\pi N \Delta}$ 
will be needed  in a combined 
analysis of the world data of $\pi N, \gamma N \rightarrow \pi N, \pi\pi N$
reactions.

We have also investigated the sensitivity of each $\gamma N\to N^\ast$
process to the $\gamma N\to \pi N$ and $\gamma N\to \pi\pi N$ reactions.
The $\gamma N\to S_{11}(1535)$, $\gamma N\to S_{31}(1620)$ and 
$\gamma N\to D_{13}(1520)$ processes are found to have significant influence
on both the single and the double pion photoproduction observables.
In particular, $\gamma N\to D_{13}(1520)$ will be key to fixing 
the overestimation at the first peak of $\gamma N\to\pi^0\pi^0 p$ around
$W=1.5$ GeV. 
As for the $P$ wave resonances, the $\gamma N\to \Delta (1232)$ process
is critical for describing the $\gamma N\to\pi N$ observables up
to $W=1.5$ GeV, while it plays almost no role for the 
total cross sections and invariant mass distributions
of $\gamma N\to\pi\pi N$ reactions.
The $\gamma N\to N^\ast (1440)$ process just has a negligible contribution
to the $\gamma N\to\pi N, \pi\pi N$ observables considered in this paper. 
This result for the $N^\ast(1440)$ is consistent with the recent analysis
in Ref.~\cite{elsa}.
The $N^*$ states that are found to be important 
in determining the single and double photoproduction reactions 
are indicated in the second and third columns of Table~\ref{tab1}.

The results in this paper show clearly that in general
the analysis of the single pion production reactions is not enough
to pin down the amplitudes associated with the electromagnetic interactions.
To extract the reliable information on the $N^\ast$ states below
$W=2$ GeV, at least one needs to perform simultaneous analysis of
the single and double pion production reactions.
Currently, this is one of the main efforts at EBAC.

\begin{acknowledgments}
The authors would like to thank Dr. V.~Mokeev for sending
the invariant mass distribution data from CLAS.
This work is supported by 
the U.S. Department of Energy, Office of Nuclear Physics Division, under 
Contract No. DE-AC02-06CH11357, and Contract No. DE-AC05-06OR23177 
under which Jefferson Science Associates operates Jefferson Lab,
by the Japan Society for the Promotion of Science,
Grant-in-Aid for Scientific Research(C) 20540270;
and by a CPAN Consolider INGENIO CSD 2007-0042 contract 
and Grant No. FIS2008-1661 (Spain).
This work used resources of the National Energy Research Scientific
Computing Center, which is supported by the Office of Science of the 
U.S. Department of Energy under Contract No. DE-AC02-05CH11231.
\end{acknowledgments}

\end{document}